\documentclass[twocolumn,preprintnumbers,nofootinbib,showpacs]{revtex4-1}

\usepackage{mathrsfs}
\usepackage{amssymb,amsmath,mathtools}
\usepackage{amssymb}
\usepackage{dsfont}
\usepackage{graphicx}
\usepackage[usenames,dvipsnames,svgnames,x11names,table]{xcolor}
\usepackage{bm}
\usepackage{subfigure}
\usepackage{hyperref}
\usepackage[normalem]{ulem}
\usepackage{dcolumn}
\usepackage{blindtext}
\usepackage[utf8]{inputenc}
\usepackage{xfrac}

\DeclarePairedDelimiter\bra{\langle}{\rvert}
\DeclarePairedDelimiter\ket{\lvert}{\rangle}
\usepackage{anysize}
\usepackage{amsmath}
\usepackage{physics}
\usepackage{soul}

\graphicspath{
    {converted_graphics/}
    {Figuras/}
}

\begin{document}

\title{Quantum State Preparation of Hydrogen Atoms by Hyperfine Quenching}

\author{Amanda Alencar*, I. Prazeres, C.~R.~de~Carvalho, F. Impens,  A. Medina, N.~V.~de~Castro~Faria, J. Robert**,  Ginette Jalbert}
\let\thefootnote\relax\footnotetext{\hspace{-0.3cm} * a.alencar@pos.if.ufrj.br}
\affiliation{
Instituto de F\'\i sica, Universidade Federal do Rio de Janeiro, Cx. Postal 68528, Rio de Janeiro, RJ 21941-972, Brazil
}
\affiliation{** Université Paris-Saclay, CNRS, Laboratoire de physique des gaz et des plasmas, 91405, Orsay, France
}

\begin{abstract}
We discuss the use of a region of uniform and constant magnetic field in order to implement a two-state atomic polarizer for an H(2S) beam. We have observed that a device with such field configuration is capable of achieving an efficient polarization for a wide range of magnetic field intensities and atomic velocities. In addition, we establish a criterion that must be met to confirm a successful polarization. That is possible due to a specific beating pattern for the Lyman-$\alpha$ radiation expected for the outgoing two-state atomic beam.

\end{abstract}

\maketitle
\section{Introduction}

Atomic structures are sensitive to the presence of external fields. For instance, light shifts may significantly alter the frequencies obtained in optical atomic clock~\cite{katori2003ultrastable}. Spectral line widths are also sensitive to applied fields. Recently, it has been shown that the lifetime of hydrogen metastable state can be influenced not only by the field intensity, but also by the chosen geometry of the electromagnetic fields~\cite{trappe2016geometric}. This paper focuses on the use of external fields to filter specific atomic states. 

The atomic polarizer, which filters specific atomic states, is an essential component of matter-wave interferometers based on a multiple-state atomic source~\cite{miniatura1992,Kouchi2019entangled} 
such as hydrogen Stern-Gerlach interferometers~\cite{miniaturaJPhysII91,robertJPhysII92,Impens17}. In the present work, we are interested in exploring the filtering of two specific metastable hyperfine structure states of the hydrogen atom. An implementation of this particular atomic polarizer can be obtained with of a region of static and uniform magnetic field. The role of the polarizer's magnetic field is twofold: it is used to tune the energy levels of the metastable hydrogen and also to generate the commoving electric field responsible for the decay, as shown in the classical references~\cite{Lamb50,Lamb52a}. The couplings between $2S_{1/2}$ and $2P_{1/2}$ atomic states dressed by the magnetic field filter out two specific $2S_{1/2}$ hyperfine structure states. We analyse the Lyman-$\alpha$ radiation rate associated with these decays and show that a specific pattern is obtained. This pattern by itself can be seen as an indication of a successful polarization process. However, a more robust pattern can also be observed if, after the desired polarization, we add an external electric field and observe the resulting Lyman-$\alpha$ emission, this time coming from the states transmitted by the initial polarization stage.   

The paper is organized as follows. Section~\ref{Td} reviews the influence of the magnetic field on the $2S_{1/2}$ and $2P_{1/2}$ hyperfine structure states of the hydrogen atom. We then derive the time-dependent probabilities to find the hydrogen atom in a specific $2S_{1/2}$ or $2P_{1/2}$ hyperfine structure state. These expressions are used in Section~\ref{sec:fast} to propose a polarizing device consisting in a region of uniform and constant magnetic field. Finally, we discuss a criterion of effectiveness for the polarizer based on the Lyman-$\alpha$ radiation patterns produced by the decay of the atomic states of interest.

\section{Theoretical description}
\label{Td}

\subsection{Zeeman effect in hyperfine $H$ states}

We review here the Zeeman effect in the hyperfine structure states $2S_{1/2}$ and $2P_{1/2}$ of the hydrogen atom. As usual, we use decomposition on a basis of orbital angular momentum, electron spin, and nucleus spin eigenstates (Appendix~\ref{apendA}) to analyse the Zeeman effect. For this purpose, we use a decomposition on the hyperfine eigenstates $\ket{2L_{J},F,M_F}$, where $\vec{F}=\vec{J}+\vec{I}=(\vec{L}+\vec{S})+\vec{I},$ with $\vec{L}$ standing for the orbital angular momentum, $\vec{S}$ the electron spin and $\vec{I}$ the spin of the nucleus. We consider the following hyperfine structure states: 
\begin{eqnarray}
&\ket{2S_{\frac{1}{2}},0,0},\ket{2S_{\frac{1}{2}},1,-1},\ket{2S_{\frac{1}{2}},1,0},\ket{2S_{\frac{1}{2}},1,1}, \nonumber \\
&\ket{2P_{\frac{1}{2}},0,0},\ket{2P_{\frac{1}{2}},1,-1},\ket{2P_{\frac{1}{2}},1,0},\ket{2P_{\frac{1}{2}},1,1}. \nonumber
\end{eqnarray}

The $2P_{3/2}$ states were not considered in this study for being about ten times further from the $2S_{1/2}$ states, regarding energy intervals. For that reason, the influence of the $\ket{2P_{3/2},F,M_F}$ states in the $\ket{2S_{1/2},F,M_F}$ states lifetimes is much less significant than the influence of the $\ket{2P_{1/2},F,M_F}$ states.

We will consider the total Hamiltonian operator
$
\hat{H}=\hat{H}_0+\hat{H}_B.
$ 
 $\hat{H}_0$ is the non-perturbed Hamiltonian expressed in terms of hyperfine structure states~\cite{Gasenzer12metastable}. $\hat{H}_B=(\hat{L}_z+g_e\hat{S}_{ez}-\tilde{g}_p\hat{S}_{pz})\mu_BB/\hbar$ describes the atom-magnetic field interaction, where $\mu_{B}$ is the Bohr magneton, $g_e$ is the gyromagnetic factor of the electron, $\tilde{g}_p= g_p \mu_N /\mu_B=g_p m_e / m_p$ is the gyromagnetic factor of the nucleus multiplied by the ratio between the electron mass and the proton mass.

The eigenvalues of the Hamiltonian $\hat{H}$ in the subspace spanned by the $\ket{2S_{1/2},F,M_F}$ states can be easily determined. The same can be done for the $\ket{2P_{1/2},F,M_F}$ states. The expressions for the mentioned Hamiltonian eigenvalues are displayed in Appendix~\ref{apendB} and Fig.~\ref{Fig1} shows them as a function of the magnetic field intensity. We take as reference the energy of the fine state $2P_{1/2}$ in the absence of magnetic field. 

\begin{figure}[h]
\centering
\textit{}\includegraphics[scale=0.38]{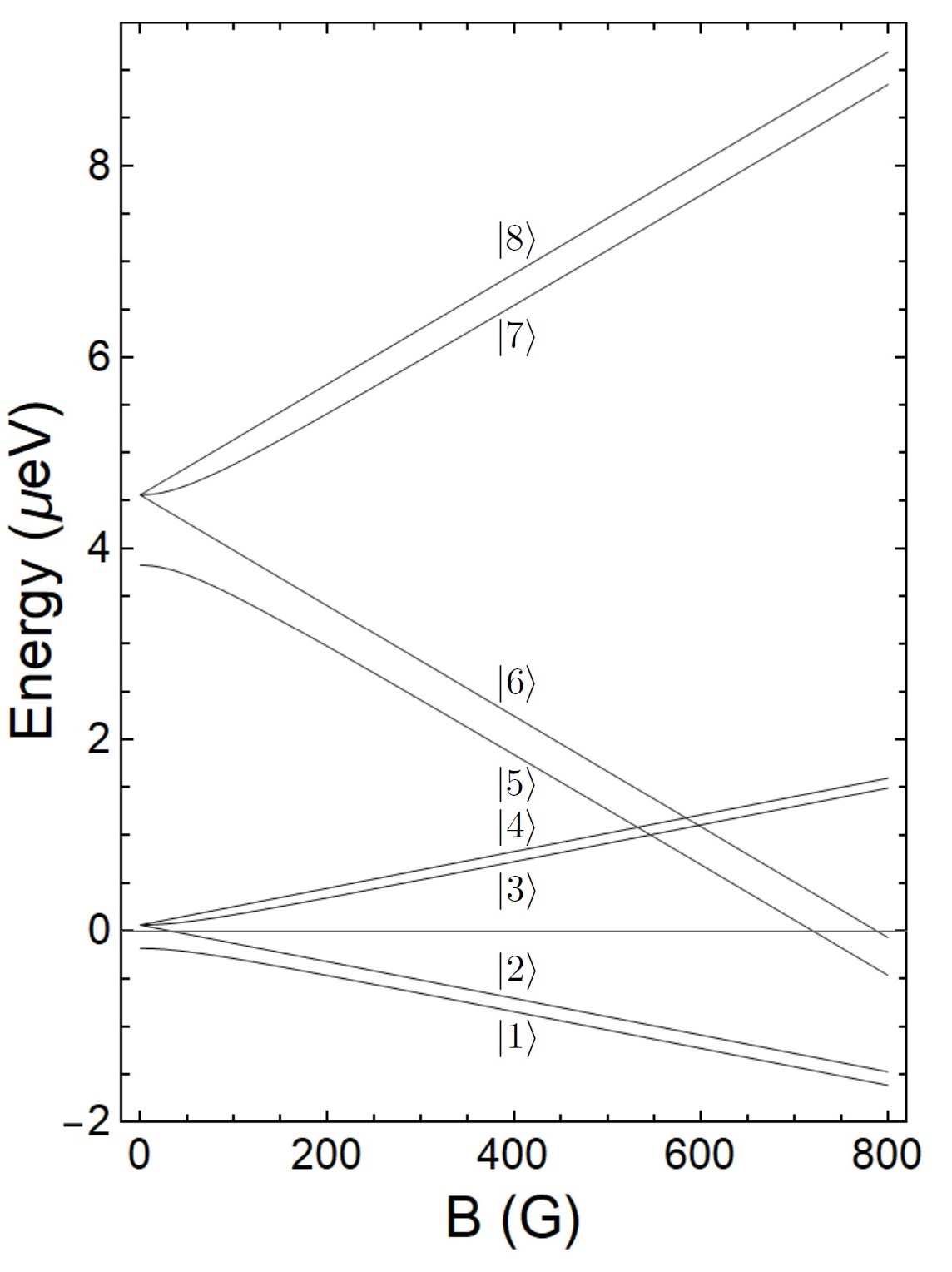}
\caption {Zeeman splitting of the $2S_{1/2}$ and $2P_{1/2}$ hyperfine structure states. The energy eigenvalues represented are $h_1$ to $h_8$ for increasing energy at low magnetic field intensities.}
\label{Fig1}
\end{figure}

As expected, the energy eigenvalues associated to the angular momentum projections $M_F= \pm 1$, Eqs. (\ref{h2},\ref{h4},\ref{h6},\ref{h8}), behave linearly with the magnetic field intensity, while the energy eigenvalues associated to the angular momentum projections $M_F=0$, Eqs. (\ref{h1},\ref{h3},\ref{h5},\ref{h7}), behave quadratically with the magnetic field intensity when $(g_e+g_p)^{2} \mu_B ^2 B^2 \ll A_1 ^2$ and $(4-g_e+3g_p)^2\mu_B^2B^2 \ll 9A_2^2$, respectively, or $B \ll 63 $G. This behaviour is related to the mixture between the states $\ket{2S_{1/2},0,0}$ and $\ket{2S_{1/2},1,0}$ and between the states $\ket{2P_{1/2},0,0}$ and $\ket{2P_{1/2},1,0}$ caused by the external magnetic field. The new states can be written with respect to the $\ket{2S_{1/2},F,M_F}$ and $\ket{2P_{1/2},F,M_F}$ states as follows:

\begin{equation}\label{1}
\ket{1}=m_1(B) \ket{2P_{\frac{1}{2}},0,0} +n_1(B) \ket{2P_{\frac{1}{2}},1,0}
\end{equation}
\begin{equation}\label{2}
\ket{2}=\ket{2P_{\frac{1}{2}},1,-1}
\end{equation}
\begin{equation}\label{3}
\ket{3}=m_3(B) \ket{2P_{\frac{1}{2}},0,0} +n_3(B) \ket{2P_{\frac{1}{2}},1,0}
\end{equation}
\begin{equation}\label{4}
\ket{4}=\ket{2P_{\frac{1}{2}},1,1}
\end{equation}
\begin{equation}\label{5}
\ket{5}=m_5(B) \ket{2S_{\frac{1}{2}},0,0} +n_5(B) \ket{2S_{\frac{1}{2}},1,0}
\end{equation}
\begin{equation}\label{6}
\ket{6}=\ket{2S_{\frac{1}{2}},1,-1}
\end{equation}
\begin{equation}\label{7}
\ket{7}=m_7(B) \ket{2S_{\frac{1}{2}},0,0} +n_7(B) \ket{2S_{\frac{1}{2}},1,0}
\end{equation}
\begin{equation}\label{8}
\ket{8}=\ket{2S_{\frac{1}{2}},1,1}
\end{equation}

In the absence of an external magnetic field, we have $n_1(0)=m_3(0)=n_5(0)=m_7(0)=0$ while  $m_1(0)=n_3(0)=m_5(0)=n_7(0)=1$. That means that all states are given in terms of only one state $\ket{2S_{1/2},F,M_F}$ or $\ket{2P_{1/2},F,M_F}$. For strong magnetic fields, on the other hand, we have $m_1(B)=n_1(B)=n_3(B)=m_5(B)=n_5(B)=n_7(B)\rightarrow 1/\sqrt{2}$, while $m_3(B)=m_7(B)\rightarrow -1/\sqrt{2}$. This means that, for high values of the magnetic field intensity, the states $\ket{5}$ and $\ket{7}$ are perfect mixtures of the states $\ket{2S_{1/2},0,0}$ and $\ket{2S_{1/2},1,0}$ and the states $\ket{1}$ and $\ket{3}$ are perfect mixtures of the states $\ket{2P_{1/2},0,0}$ and $\ket{2P_{1/2},1,0}$.

Figure~\ref{Fig1} reveals crossings between some of the $2S_{1/2}$ and $2P_{1/2}$ levels for well-defined values of the magnetic field. A strong coupling between certain states can be achieved by the motional electric field in the vicinity of the crossing regions. Selection rules impose $\Delta m_l=\pm 1$  with the motional electric field perpendicular to the quantization direction (See Appendix~\ref{apendC}).
   
Since the $2P_{1/2}$ states have short lifetime, of the order of $10^{-9}$s, it is possible to induce the decay of the $2S_{1/2}$ metastable states by creating a channel to the ground state through the unstable $2P_{1/2}$ states. Even a small external magnetic field may significantly enhance the decay probability of a moving atom in a $2S_{1/2}$ state. We study below how external fields affect the lifetimes of hyperfine states.

\subsection{Time-dependent evolution of the atomic states}

The quenching process will be described using time dependent perturbation theory taking $\gamma_{2s}\approx 7 s^{-1}$ and $\gamma_{2p}\approx 6\times 10^{8} s^{-1}$ as the $2S_{1/2}$ and $2P_{1/2}$ decay constants, noting that $\gamma_{2s} \ll \gamma_{2p}$. The expressions below give the temporal variation of the probability amplitude of the atom to be observed in a certain state considering external perturbation and spontaneous decay
\begin{equation}\label{ddtcsdet}
\frac{d}{dt}c_j(t)=-\frac{1}{2}\gamma_{2s}c_j(t) + \sum_{k \in P} \frac{V_{jk}}{i\hbar}e^{i\omega_{jk} t}c_k(t)
\end{equation} 
\begin{equation}\label{ddtcpdet}
\frac{d}{dt}c_k(t)=-\frac{1}{2}\gamma_{2p}c_k(t) + \sum_{j \in S} \frac{V_{kj}}{i\hbar}e^{i\omega_{kj} t}c_j(t)
\end{equation}
where the sets $P = \{1,2,3,4 \} $ and  $S  = \{ 5,6,7,8 \} $ corresponds to the $2P_{1/2}$ and $2S_{1/2}$ states, respectively. The indices $j$ and $k$ denote $2S_{1/2}$ and $2P_{1/2}$ states, respectively.  $c_j(t)$ and $c_k(t)$ are the corresponding time-dependent probability amplitudes. As discussed later, each state of the set $S$($P$) is coupled to two states of the set $P$($S$). However, as discussed in Appendix~\ref{apendC}, only one of these two couplings will be relevant for the quantum evolution. The relevant coupling depends on the electric field induced in the atomic frame as well as on the magnetic field.

Since these states are separated by an energy $\Delta \varepsilon_{jk} = h_j-h_k = \hbar \omega_{jk} $, being those states coupled by an electric field $\vec{E}$, the matrix elements $V_{kj}$ are:
\begin{equation}
V_{kj}=\bra{k}e\vec{E} \cdot \vec{r} \ket{j}=V_{jk}^*.
\end{equation}  
See Appendix~\ref{apendC}.

The solutions $c_j(t)$ and $c_k(t)$ can be found by decoupling (\ref{ddtcsdet}) and (\ref{ddtcpdet}) through differentiation.

We take the atom initially in the $2S_{1/2}$ level so that $ c_j(0)=1 $ and $ c_k(0)=0 $.
The solutions can then be written as:
{\fontsize{9.5}{13}\selectfont
\begin{equation}\label{csdet}
c_j(t)=\Big( \frac{\lambda_{2}^{+}+\frac{1}{2}\gamma_{2s}}{\lambda_{2}^{+}-\lambda_{1}^{+}} \Big) e^{\lambda_{1}^{+} t}-\Big( \frac{\lambda_{1}^{+}+\frac{1}{2}\gamma_{2s}}{\lambda_{2}^{+}-\lambda_{1}^{+}} \Big)e^{\lambda_{2}^{+} t}
\end{equation}
\begin{equation}\label{cpdet}
c_k(t)=-\Big( \frac{(i\hbar)^{-1} V_{kj}}{\lambda_{2}^{-}-\lambda_{1}^{-}} \Big)e^{\lambda_{1}^{-} t}+\Big( \frac{(i\hbar)^{-1} V_{kj}}{\lambda_{2}^{-}-\lambda_{1}^{-}} \Big)e^{\lambda_{2}^{-} t}
\end{equation}
}
with $\lambda_{1,2}^{+}= -\gamma_{+}/2 \pm \sqrt{\gamma_{+}^2/4-\omega_{0_+}^2}$ and $\lambda_{1,2}^{-}= -\gamma_{-}/2 \pm \sqrt{\gamma_{-}^2/4
-\omega_{0_-}^2}$
and where $\gamma_{\pm} = (\gamma_{2s}+\gamma_{2p})/2 \pm i\omega_{kj}$, $\omega_{0_+}^2= |V_{kj}|^2/\hbar^2+\gamma_{2s}\gamma_{2p}/4 + i\gamma_{2s}\omega_{kj}/2$ and $\omega_{0_-}^2= |V_{kj}|^2/\hbar^2+\gamma_{2s}\gamma_{2p}/4 - i\gamma_{2p}\omega_{kj}/2$. 
$\gamma_{\pm}$ and $\omega_{0_\pm}$ are dependent of the energy difference $\Delta \varepsilon_{jk}$, but that dependence is not explicitly expressed for the sake of simplicity.

The considered probabilities are then:
\begin{equation}\label{probtrabalhada}
\begin{aligned}
&|c_j(t)|^2 = \hspace{0.13cm} \frac{e^{-\Re(\gamma_+)t}}{\abs{2\beta_+}^2} \hspace{0.13cm} \Bigg\{ {} \bigg \{ \frac{\gamma_{2s}^2}{4}-\gamma_{2s} \Re\Big( \frac{\gamma_{+}}{2}+ \beta_+ \Big) \\
	 &+ \abs{ \frac{\gamma_{+}}{2}+ \beta_+ } ^2 \bigg \}\hspace{-0.03cm} e^{2\Re ( \beta_+ ) t} \hspace{-0.07cm} + \hspace{-0.08cm} \bigg \{\hspace{-0.07cm} \frac{\gamma_{2s}^2}{4}\hspace{-0.07cm} -\hspace{-0.05cm} \gamma_{2s} \hspace{-0.04cm} \Re\hspace{-0.03cm}\Big(\hspace{-0.03cm} \frac{\gamma_{+}}{2}\hspace{-0.07cm}-\hspace{-0.07cm} \beta_+ \hspace{-0.05cm} \Big) \\
    & +\abs{ \frac{\gamma_{+}}{2}- \beta_+ } ^2 \bigg \}  e^{-2\Re ( \beta_+ ) t} - \hspace{-0.02cm} \bigg \{ \frac{\gamma_{2s}^2}{4}-\gamma_{2s}\Re\Big( \frac{\gamma_{+}}{2} \Big)  \\
     & + \abs{\frac{\gamma_{+}}{2}}^2 -\abs{\beta_+}^2 \bigg \} \Big( e^{2i\Im ( \beta_+ ) t}+e^{-2i\Im ( \beta_+ ) t} \Big)\\
     & -  2i {} \bigg \{ \Re\Big( \frac{\gamma_{+}}{2} \Big)\Im ( \beta_+ ) -\Im\Big( \frac{\gamma_{+}}{2} \Big)\Re\big( \beta_+ \big)  \\
     &  -\frac{1}{2}\gamma_{2s}\Im\big( \beta_+ \big)\bigg \} \Big( e^{2i\Im ( \beta_+ ) t} - e^{-2i\Im ( \beta_+ ) t} \Big) \Bigg \},
\end{aligned}
\end{equation}
and
\begin{equation}\label{modck2}
\begin{aligned}
|c_k(t)|^2 & = \frac{|V_{kj}|^2 e^{-\Re(\gamma_-)t}}{\hbar^2 \abs{2\beta_-}^2} \hspace{0.13cm} \Bigg\{ {} e^{2\Re ( \beta_- ) t} +  \\
     &  + e^{-2\Re ( \beta_- ) t} -2 \cos \big( 2 \Im ( \beta_- ) t \big) \Bigg \}
\end{aligned}
\end{equation}
where $\beta_{\pm}=\sqrt{\gamma^2_{\pm}/4-\omega_{0\pm
}^2}$.

Eq.~(\ref{probtrabalhada}), although seemingly complicated, falls for small values of the electric field, $|V_{kj}|^2 \ll \hbar^2 (\gamma_{2p}^2+4 \omega_{kj}^2)$, into the well-known expression \cite{Lamb50}
\begin{equation}\label{probapp1}
|c_j(t)|^2 \approx e^{-\gamma_{E}t},
\end{equation}
where $\gamma_{E}=\gamma_{2p}|V_{kj}|^2/\big(\hbar^2 (\gamma_{2p}^2/4 + \omega_{kj}^2 )\big)$ is the resulting decay rate of the state $2S_{1/2}$ due to the action of an external electric field.

\subsection{Decay for a resonant coupling}

When the energy difference between the two coupled states cancels, for a specific magnetic field intensity, a simpler interpretation of Eq.~(\ref{probtrabalhada}) is possible. With this condition, the parameters $\omega_{0+}$ and $\gamma_+$ turn real and can be interpreted as a frequency of oscillation and the damping constant, respectively, of a damped harmonic oscillator. Thus,
$\gamma_{\pm}=\gamma=(\gamma_{2p}+\gamma_{2s})/2$
and
$\omega_{0\pm}^2=\omega_0^2=|V_{kj}|^2/\hbar^2+(\gamma_{2s}\gamma_{2p})/4.$

Besides, Eq.~(\ref{probtrabalhada}) presents three different regimes depending on whether $\gamma ^2/4<\omega_0^2$, $\gamma ^2/4>\omega_0^2$ or $\gamma ^2/4=\omega_0^2$, corresponding to the under-damped, over-damped and critically damped regimes, respectively.

In the under-damped regime, we can write Eq.~(\ref{probtrabalhada}) as:
\begin{equation}\label{sub}
\begin{aligned}
|c_j(t)|^2= \bigg \{  & \frac{\bar{\gamma}^2}{4} + \omega^2  - \bigg ( \frac{\bar{\gamma}^2}{4} - \omega^2 \bigg )\cos(2\omega t)\\
& +\bar{\gamma} \omega \sin(2\omega t) \bigg \} \frac{e^{-\gamma t}}{2\omega^2},
\end{aligned}
\end{equation}
where $\bar{\gamma}=(\gamma_{2p}-\gamma_{2s})/2$ and $\omega=\sqrt{\omega_{0
}^2-\gamma^2/4}$.

For a sufficiently strong damping, meeting the condition $\gamma ^2/4>\omega_0^2$, condition in which the decay happens without oscillations, we have that:
\begin{equation}\label{super}
\begin{aligned}
|c_j(t)|^2= \bigg \{ \Big( & \beta + \frac{\bar{\gamma}}{2}\Big )^2 e^{2\beta t} + \Big(\beta - \frac{\bar{\gamma}}{2}\Big )^2 e^{-2\beta t}   \\
& + 2\beta^2 -\frac{1}{2} \bar{\gamma}^2 \bigg \} \frac{e^{-\gamma t}}{4\beta^2},
\end{aligned}
\end{equation}
where $\beta=\sqrt{\gamma^2/4-\omega_{0
}^2}$.

In the critically damped regime, we can write:
\begin{equation}\label{critico}
|c_j(t)|^2= \bigg \{ 1 + \bar{\gamma} + \frac{1}{4}\bar{\gamma}^2 t^2 \bigg \} e^{-\gamma t}.
\end{equation} 

The three different regimes of decay can be achieved by varying the atom's speed or with the addition of an external electric field.

\section{Results and Discussion}
\label{sec:fast}

In this Section, we apply the previous analysis of the Zeeman energy levels to discuss the working principle of an atomic polarizer. We also analyse the Lyman radiation rate emitted by the selected atomic fragments. This rate exhibits a characteristic pattern.

\subsection{The atomic polarizer}

\begin{figure}[h]
\includegraphics[scale=.34]{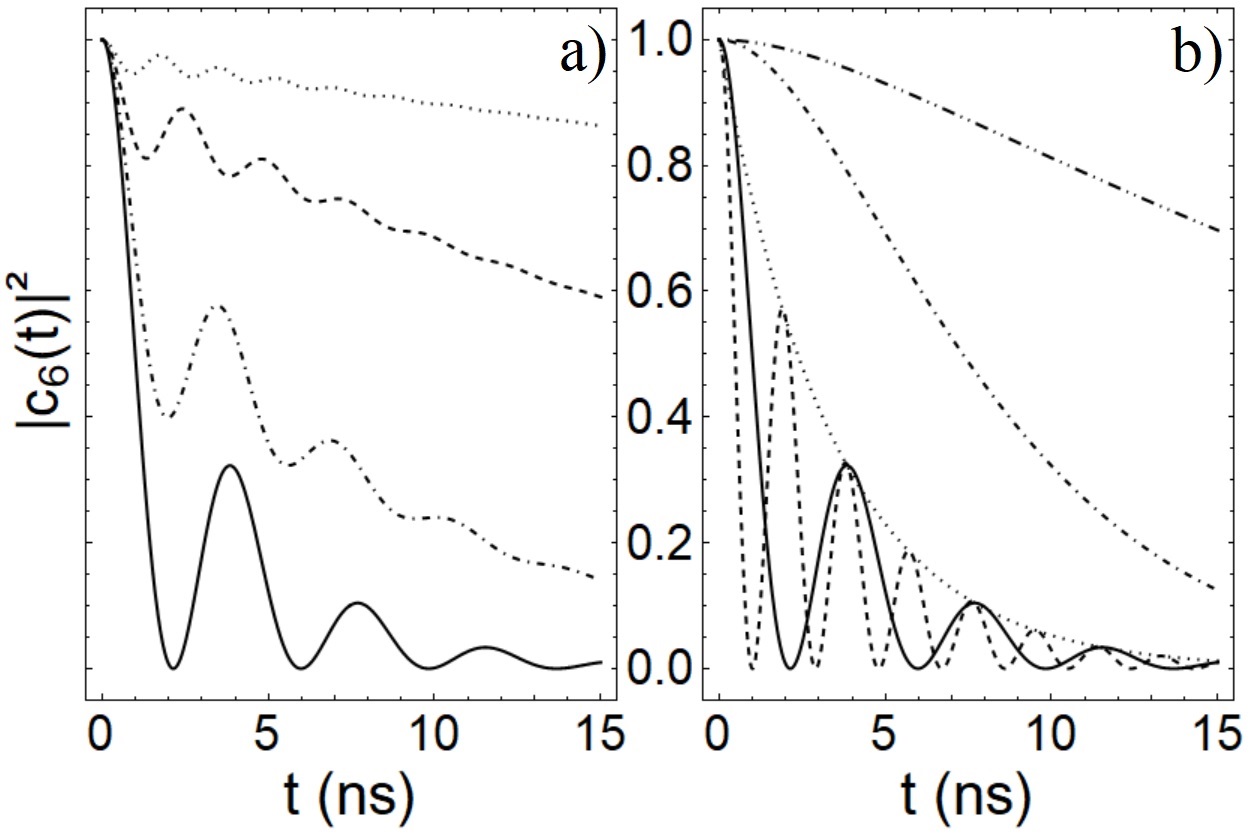}
\caption{a) Probability, $|c_{6}(t)|^2$, for a magnetic field intensity of $300 \: {\rm G}$ (dotted line), $400 \: {\rm G}$ (dashed line), $500 \: {\rm G}$ (dot-dashed line) and $B_{3,6}= 597 \: {\rm G}$ (solid line) for an atomic speed $v \: = \: 100 \: {\rm km/s}$. 
b) Probability $|c_{6}(t)|^2$ for the magnetic field $B_{3,6}$ and with the atomic velocities: 100 km/s (solid line) and 200 km/s (dashed line), corresponding to the underdamped regime; 17.7 km/s (dot-dashed), corresponding to the critically damped regime; and 8 km/s (dot-dot-dashed), corresponding to the overdamped regime. 
The dotted line shows the envelope $e^{-\frac{1}{2}(\gamma_{2s}+\gamma_{2p})t}$.}
\label{Fig2}
\end{figure}

We consider the decay rate of $2 S_{1/2}$ atoms (state $|6 \rangle =\ket{2S_{1/2},1,-1}$) propagating in a region of constant magnetic field with a velocity orthogonal to the field. In particular, we investigate the behaviour of this decay in the vicinity of the magnetic field $B_{3,6}=597 \: {\rm G}$ corresponding to the crossing between states $|3\rangle$ and $|6\rangle$. The probability of the atom being in the corresponding $2 S_{1/2}$ state as a function of time (given by Eq.~(\ref{probtrabalhada})) is presented on Figure~\ref{Fig2}. The relevant coupling in this region occurs between the states $|3\rangle$ and $|6\rangle$. Indeed, the couplings between $|6\rangle$ and $|4\rangle$ as well as $|6\rangle$ and $|2\rangle$ are forbidden by the selection rules. On the other hand, the coupling between $|6\rangle$ and $|1\rangle$ is strongly non-resonant and also suppressed by the asymptotic form of the eigenstate $|1 \rangle$ for strong magnetic field values (as detailed in Appendix~\ref{apendC}).
   
Fig.~\ref{Fig2}a) shows that the decay depends strongly on the magnetic field intensity and is enhanced in the vicinity of the crossing value $B_{3,6}$. Fig.~\ref{Fig2}b) displays the time-dependent probability in the different regimes, Eqs.(~\ref{sub}-\ref{critico}), at the crossing magnetic field $B_{3,6}$ for four atomic velocities, with $\vec{v} \perp \vec{B}$. In the underdamped regime, the probability of decay has an exponential envelope $e^{-\frac{1}{2}(\gamma_{2s}+\gamma_{2p})t}$. This can be interpreted as follows: the atom undergoes Rabi oscillations between states $\ket{6}$ and $\ket{3}$ and spends on average an equal time in both states. Since one of these states ($2S_{1/2}$) has a much smaller decay rate, the lifetime of the moving atom is approximately twice that of the $2P_{1/2}$ state. The maximum decay is obtained at the crossing magnetic field value $B_{3,6}$. 

The discussion above can be extended to the state $\ket{5}$ (coupled to state $\ket{4}$). The remaining $2 S_{1/2}$ states $(\ket{7},\ket{8})$ do not exhibit energy crossings and are thus not resonantly coupled. Indeed, their lifetimes (less affected by the presence of the magnetic field) are orders of magnitude larger than that of the states $\ket{5}$ and $\ket{6}$.  This fact enables one to build a device which quenches two of the four $2S_{1/2}$ Zeeman states, i.e., which filters out the states $\ket{5}$ and $\ket{6}$ and preserves the states $\ket{7}$ and $\ket{8}$, consisting of a region of uniform and constant magnetic field with appropriate length through which the atomic beam must travel, as further discussed in Appendix~\ref{apendD}.

\subsection{Detection signal with Lyman-$\alpha$ radiation and Criterion for testing a polarizing device}

Detection can be done by sensing the Lyman-$\alpha$ radiation emitted by the propagating fragments. Since the probability of emission of Lyman-$\alpha$ radiation is $10^8$ times higher when the atom is in a $2P_{1/2}$ state when compared to a $2S_{1/2}$ state, the radiation rate may be taken as proportional to the probability of the atom being in one of the $2P_{1/2}$ states:
\begin{equation}\label{Plyman2P}
R_{Lyman-\alpha} \propto \sum_{k=1}^{4} |c_k(t)|^2, 
\end{equation}
where the probabilities $|c_k(t)|^2$ are given by Eq.(\ref{modck2}).

Let us consider a non-polarized H(2S) beam (with four equally populated $2S_{1/2}$ Zeeman states), propagating in the x-direction through a region with an orthogonal magnetic field yielding near-resonant energy levels $\ket{3},\ket{4},\ket{5},\ket{6}$.  

\begin{figure}[h!]
\includegraphics[scale=.49]{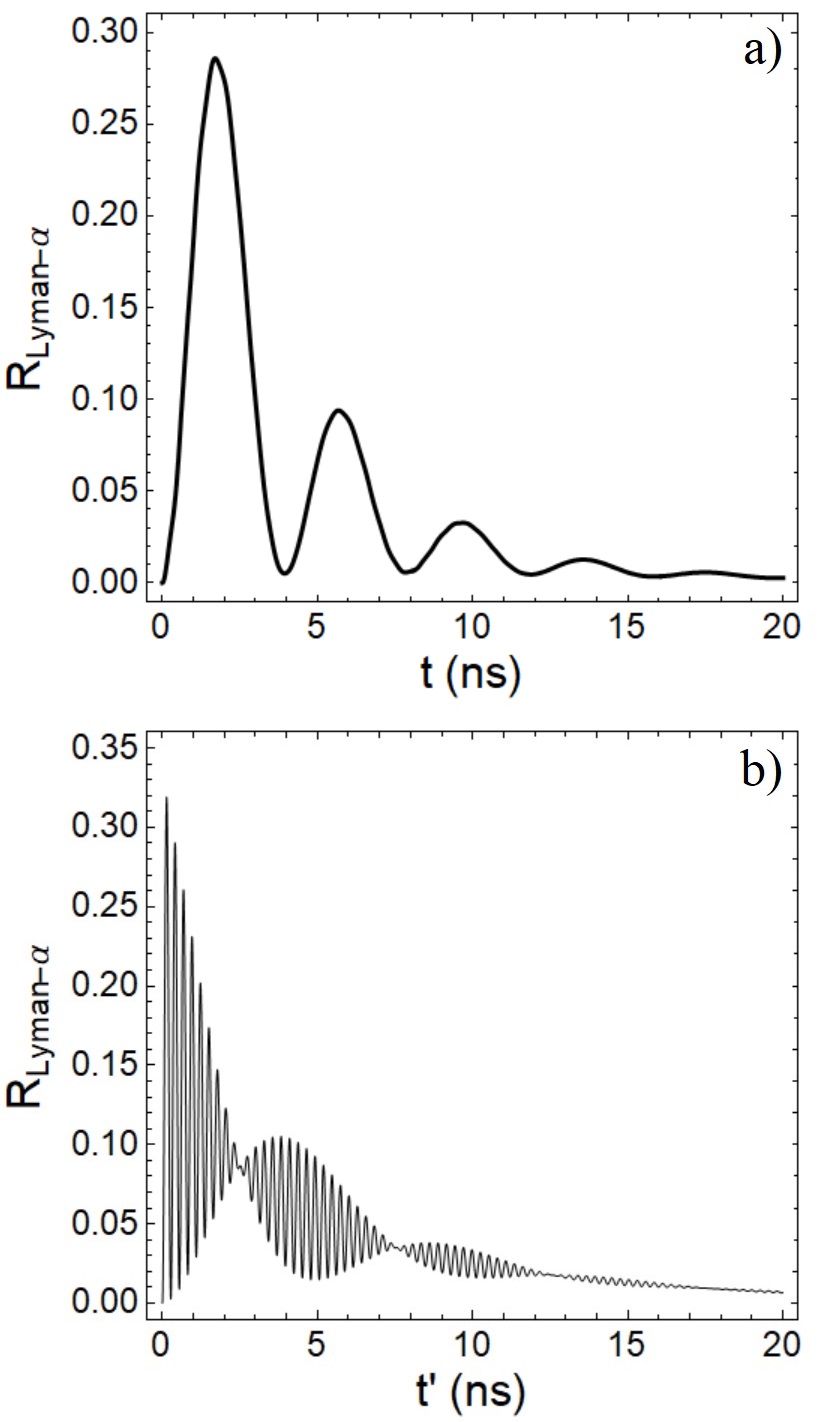}
\caption{a) Joint Lyman-$\alpha$ emission originated by the quenching of the four $2S_{1/2}$ states when a H(2S) beam of velocity $\vec{v} = v \hat{x}$ with 
 $v \: = \: 100 \: {\rm km/s}$ travels through a region with a magnetic field of $\vec{B}= B \hat{z}$  ($B= 565 G$). b) Joint Lyman-$\alpha$ emission originated by the quenching of states $\ket{7}$ and $\ket{8}$ in presence of an identical magnetic field and an additional external electric field $\vec{E}_{ext}=E_{ext} \hat{y}$ with $E_{ext}=600 \: \rm{V/cm}$.}
\label{Fig3}
\end{figure}

Figure~\ref{Fig3}a) shows the profile of a joint Lyman-$\alpha$ emission rate produced by the quenching of the four $2S_{1/2}$ Zeeman states.
The dominant contribution to the Lyman-$\alpha$ radiation pattern comes from the coupling between the states $\ket{5}$ and $\ket{4}$, and $\ket{6}$ and $\ket{3}$, respectively, while the contribution from states $\ket{7}$ and $\ket{8}$ is negligible. For a magnetic field of 565 G we have an optimal value for a clear pattern, since the oscillations in the Lyman-$\alpha$ emission rate for the states that contribute the most get in phase, but significant variations in the magnetic field intensity still result in a similar pattern.

After roughly $t \simeq 20 \: {\rm ns}$ (corresponding to a path of a few mm at the considered velocity  $v \: = \: 100 \: {\rm km/s}$), most atoms that were initially in the states $\ket{5}$ and $\ket{6}$ are in the ground state, and the resultant beam is almost completely composed by atoms in states $\ket{7}$ and $\ket{8}$, since the latter present a very low Lyman-$\alpha$ emission rate when compared to the former.

Although states $\ket{7}$ and $\ket{8}$ have a similar behaviour in the presence of external fields, regarding Lyman-$\alpha$ radiation emission, their simultaneous presence in the resulting beam drastically changes the radiation pattern in presence of an additional external electric field (fundamental to increase their decay rate). This can be seen in Figure~\ref{Fig3}b), which shows the Lyman radiation in presence of an additional external electric field for an atomic beam containing initially only equally populated
$\ket{7},\ket{8}$ states. One notes an oscillatory pattern with a beat frequency $\Delta \omega =\big( \sqrt{\Delta \varepsilon_{81}^2+4|V_{81}|^2}-\sqrt{\Delta \varepsilon_{72}^2+4|V_{72}|^2} \big) / \hbar$, that can be calculated from $\vec{v}$, $\vec{B}$ and the external electric field $\vec{E}_{ext}$.

\begin{figure}[h!]
\hspace{-0.06cm}\includegraphics[scale=.335]{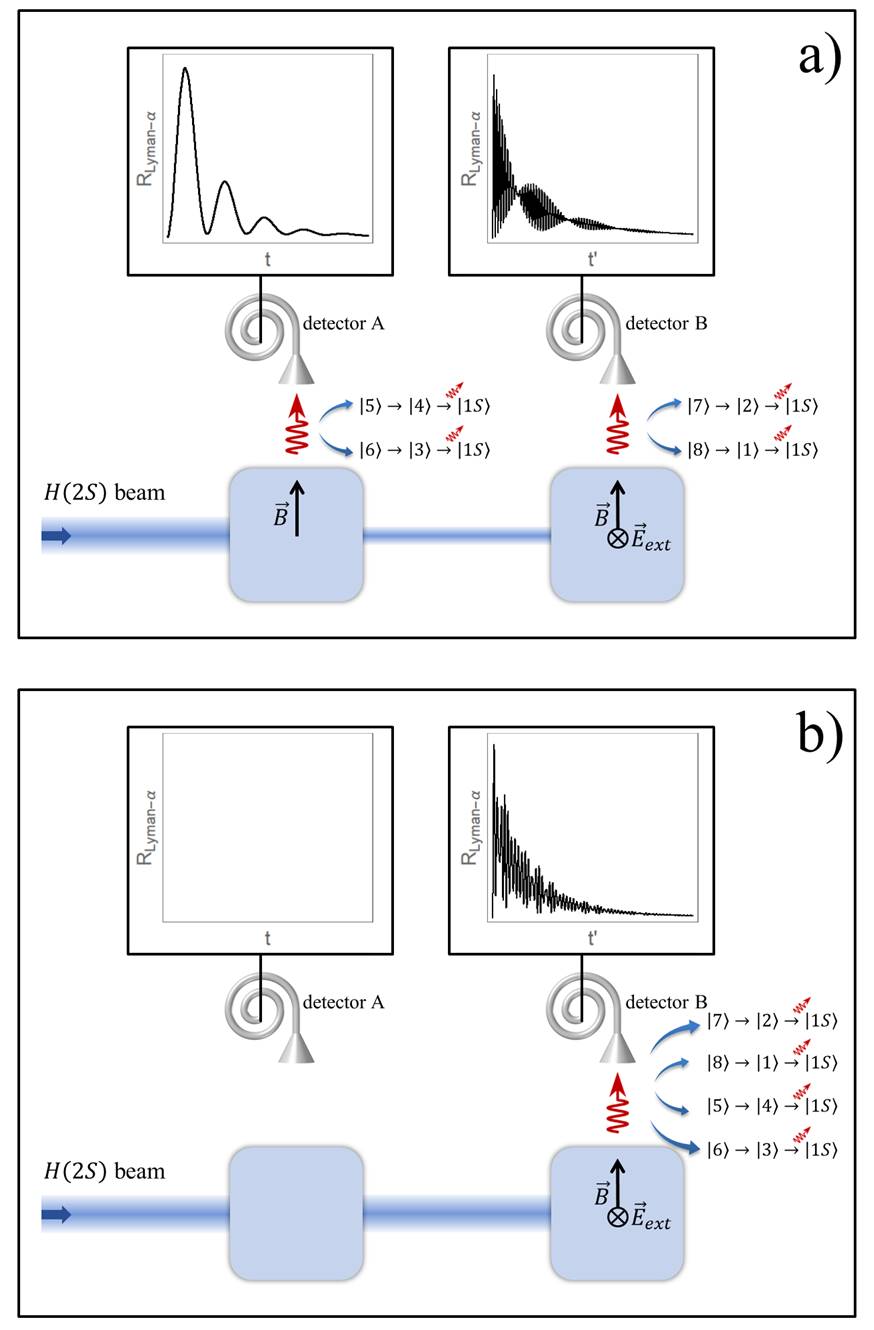}
\caption{a) Proposed setup for testing polarization with a fully working polarizer. A non-polarized H(2S) beam enters a region of constant magnetic field. In this regime, states $\ket{5}$ and $\ket{6}$ suffer quenching   and the resulting Lyman-$\alpha$ radiation is detected by detector A. Upon leaving this first region, the beam is composed exclusively of states $\ket{7}$ and $\ket{8}$. It then travels to a second region with a constant magnetic field and an additional external electric field that quenches the remaining states. The Lyman-$\alpha$ radiation is then sensed by detector B. b) Testing setup with no polarizer. A non-polarized H(2S) beam travels through the apparatus until it gets to a region of constant electric and magnetic fields. Those fields cause intense quenching of the four hyperfine structure states and the resulting Lyman-$\alpha$ radiation is detected by detector B.
The magnetic field is only present in two specific regions for illustrative purposes, it may remain constant throughout the setup.}
\label{Fig4}
\end{figure}
 
In theory, a polarizer consisting of a region of approximately constant magnetic field with adequate intensity and extension should be able to filter two of the four states of hyperfine structure, as discussed. However, inherent limitations in the apparatus construction, such as resulting inhomogeneities of the external magnetic field, can compromise the efficiency of the device. The study presented here leads to a simple and robust criterion for testing a polarizer after its construction and verify if the desired polarization has been achieved. This technique could be used prior to inserting the polarizing device into a more complex apparatus.

Figure~\ref{Fig4} shows a sketch of a setup for that verification. Fig.~\ref{Fig4}a) shows an initially non-polarized beam containing equal parts of the four hyperfine structure states. The beam then enters a region of constant magnetic field that causes quenching of two of the four states preferentially. The two surviving states then enter a region with an additional external electric field strong enough to cause a decay within a few millimetres of propagation, corresponding to a few nanoseconds. Detectors A and B should capture spectrums like the ones indicated in Fig.~\ref{Fig4}a). In the absence of a polarizer, the detection patterns would be as shown in Fig.~\ref{Fig4}b).

The pattern obtained by detector A in Fig.~\ref{Fig4}a) can be seen as an indication of a successful polarization process. However, the pattern from detector B is much more robust, as it is little affected by the atomic speed. See Appendix~\ref{apendD}. Indeed, the quenching of the two remaining states relies mostly on the external electric field, and not on the motional electric field as previously. Nevertheless, the speed still influences the detection time of the radiation, but this effect can be minimized by shortening the path and by using an atomic beam of narrow velocity distribution. The beating pattern does not depend on the specific polarizing process, as long as it delivers a beam with similar proportions of states $\ket{7}$ and $\ket{8}$.

\section{Conclusion}
\label{Conc}

In this work, we have discussed how a tuning of the metastable hydrogen energy levels by a magnetic field can enable the realization of a two-state polarizer. This is apparent from the expressions of the time-dependent probability amplitudes for the $2S_{1/2}$ and $2P_{1/2}$ states. We have particularly investigated the effects of near-resonant coupling of the metastable states dressed by the magnetic field. We have also analysed the Lyman-$\alpha$ radiation rate, used in most detection schemes, for typical quenching conditions. The simultaneous presence of two $2S_{1/2}$ states after the polarization leads to a radiation pattern with beat notes that can be seen as an indication of successful polarization. We have established a simple and robust criterion for checking the effectiveness of a magnetic polarizer in atom-interferometers setup. Perspectives for this work include the design of specific electric and magnetic field geometries enabling the hydrogen polarization along a single atomic state instead of a two-state multiplicity.

\section{acknowledgments}

This work was partially supported by the Brazilian agencies CNPq, CAPES, and FAPERJ. It is part of the INCT-IQ from CNPq.

\appendix
\section{Decomposition of $\ket{2L_J,F,M_F}$ in $\ket{L,M_L}\ket{M_e}\ket{M_p}$}
\label{apendA}
 
As usual, the decompositions of the states under study in terms of orbital angular momentum, electron spin, and nucleus spin are performed, for $L$ and $J$ fixed, as follows:
\begin{equation}
\ket{F,M_F}=\sum_{M_L,M_e,M_p} C_{L,e,p} \ket{L,M_L}\ket{M_e}\ket{M_p},
\end{equation}
where $M_e$ and $M_p$ are the projections of the electron and nucleus spin in the z axis, respectively, and can assume the values $\pm \hbar/2$, here represented as $\ket{\uparrow}$ and $\ket{\downarrow}$. The $C_{L,e,p}$ are Clebsch-Gordan coefficients. 
The results of the decompositions are the following:
{\fontsize{9.5}{13}\selectfont
$$
\begin{aligned}
&\ket{2S_{\frac{1}{2}},0,0}=-\frac{1}{\sqrt{2}}\ket{0,0}_L\hspace{-0.04cm}\ket{\uparrow}_e\hspace{-0.03cm}\ket{\downarrow}_p\hspace{-0.07cm}+\hspace{-0.06cm}\frac{1}{\sqrt{2}}\ket{0,0}_L\hspace{-0.03cm}\ket{\downarrow}_e\hspace{-0.03cm}\ket{\uparrow}_p \\ 
&\ket{2S_{\frac{1}{2}},1,-1}= \ket{0,0}_L\ket{\downarrow}_e\ket{\downarrow}_p \\
&\ket{2S_{\frac{1}{2}},1,0}= \frac{1}{\sqrt{2}}\ket{0,0}_L\ket{\uparrow}_e\ket{\downarrow}_p+\frac{1}{\sqrt{2}}\ket{0,0}_L\ket{\downarrow}_e\ket{\uparrow}_p \\
&\ket{2S_{\frac{1}{2}},1,1}= \ket{0,0}_L\ket{\uparrow}_e\ket{\uparrow}_p              \\
&\ket{2P_{\frac{1}{2}},0,0}=\frac{1}{\sqrt{3}}\ket{1,1}_L\ket{\downarrow}_e\ket{\downarrow}_p-\frac{1}{\sqrt{6}}\ket{1,0}_L\ket{\uparrow}_e\ket{\downarrow}_p \\
& \hspace{1.37cm} -\frac{1}{\sqrt{6}}\ket{1,0}_L\ket{\downarrow}_e\ket{\uparrow}_p+\frac{1}{\sqrt{3}}\ket{1,-1}_L\ket{\uparrow}_e\ket{\uparrow}_p       \\
&\ket{2P_{\frac{1}{2}},1,-1}\hspace{-0.06cm}=\hspace{-0.07cm}\sqrt{\frac{2}{3}}\hspace{-0.05cm}\ket{1,-1}_L\hspace{-0.05cm}\ket{\uparrow}_e\hspace{-0.05cm}\ket{\downarrow}_p \hspace{-0.07cm}-\hspace{-0.05cm}\frac{1}{\sqrt{3}}\hspace{-0.05cm}\ket{1,0}_L\hspace{-0.05cm}\ket{\downarrow}_e\hspace{-0.05cm}\ket{\downarrow}_p\\ 
&\ket{2P_{\frac{1}{2}},1,0}\hspace{-0.05cm}=\hspace{-0.06cm}-\frac{1}{\sqrt{3}}\ket{1,1}_L\ket{\downarrow}_e\ket{\downarrow}_p\hspace{-0.06cm}+\hspace{-0.06cm}\frac{1}{\sqrt{6}}\ket{1,0}_L\ket{\uparrow}_e\ket{\downarrow}_p \\
& \hspace{1.37cm} -\frac{1}{\sqrt{6}}\ket{1,0}_L\ket{\downarrow}_e\ket{\uparrow}_p+\frac{1}{\sqrt{3}}\ket{1,-1}_L\ket{\uparrow}_e\ket{\uparrow}_p         \\
&\ket{2P_{\frac{1}{2}},1,1}\hspace{-0.05cm}=\hspace{-0.06cm}-\sqrt{\frac{2}{3}}\ket{1,1}_L\hspace{-0.03cm}\ket{\downarrow}_e\ket{\uparrow}_p\hspace{-0.06cm}+\hspace{-0.06cm}\frac{1}{\sqrt{3}}\ket{1,0}_L\hspace{-0.03cm}\ket{\uparrow}_e\ket{\uparrow}_p
\end{aligned}
$$}

\section{Hamiltonian eigenvalues}
\label{apendB}
The equations bellow express the energy eigenvalues of the $2S_{1/2}$ and $2P_{1/2}$ Zeeman states. 

\begin{equation}\label{h1}
h_1=-\frac{A_2}{4}-\frac{1}{6} \sqrt{ 9A_2^2 +\big( 4-g_e+3g_p \big)^2\mu_B^2B^2 } 
\end{equation}
\begin{equation}\label{h2}
h_2=\frac{A_2}{4}-\frac{1}{6}(4-g_e-3g_p)\mu_BB
\end{equation}
\begin{equation}\label{h3}
h_3=-\frac{A_2}{4}+\frac{1}{6} \sqrt{ 9A_2^2 +\big( 4-g_e+3g_p \big)^2\mu_B^2B^2 }
\end{equation}
\begin{equation}\label{h4}
h_4=\frac{A_2}{4}+\frac{1}{6}(4-g_e-3g_p)\mu_BB
\end{equation}

\begin{equation}\label{h5}
h_5=-\frac{A_1}{4}+L-\frac{1}{2}\sqrt{A_1^2+(g_e+g_p)^2\mu_B^2B^2}
\end{equation}
\begin{equation}\label{h6}
h_6=\frac{A_1}{4}+L-\frac{1}{2}(g_e-g_p)\mu_BB
\end{equation}
\begin{equation}\label{h7}
h_7=-\frac{A_1}{4}+L+\frac{1}{2}\sqrt{A_1^2+(g_e+g_p)^2\mu_B^2B^2}
\end{equation}
\begin{equation}\label{h8}
h_8=\frac{A_1}{4}+L+\frac{1}{2}(g_e-g_p)\mu_BB
\end{equation}

\section{Coupling between the states $\ket{k}$ and $\ket{j}$}
\label{apendC}

The states $\ket{j}$ and $\ket{k}$ can be written in terms of the states $\ket{2S_{1/2},F,M_F}$ and $\ket{2P_{1/2},F,M_F}$, respectively, as shown by Eqs.~(\ref{1}-\ref{8}). Therefore, to determine the coupling between states $\ket{j}$ and $\ket{k}$ it is worth analysing the matrix elements $\bra{2P_{1/2},F,M_F}e \vec{E} \cdot \vec{r}\ket{2S_{1/2},F,M_F}$.

Considering the electric field in the $y$-direction in our particular case of study (the velocity of the atoms was considered to be in the $x$-direction and the magnetic field in the $z$-direction), the matrix elements of interest can be calculated. The non-zero matrix elements obtained were:
\begin{equation}\label{v16}
\bra{2P_{\frac{1}{2}},0,0}e \vec{E} \cdot \vec{r}\ket{2S_{\frac{1}{2}},1,-1}=-\frac{\sqrt{6}}{2}a_0ieE
\end{equation}
\begin{equation}\label{v36}
\bra{2P_{\frac{1}{2}},1,0}e \vec{E} \cdot \vec{r}\ket{2S_{\frac{1}{2}},1,-1}=\frac{\sqrt{6}}{2}a_0ieE
\end{equation}
\begin{equation}\label{v25}
\bra{2P_{\frac{1}{2}},1,-1}e \vec{E} \cdot \vec{r}\ket{2S_{\frac{1}{2}},0,0}=\frac{\sqrt{6}}{2}a_0ieE
\end{equation}
\begin{equation}\label{v27}
\bra{2P_{\frac{1}{2}},1,-1}e \vec{E} \cdot \vec{r}\ket{2S_{\frac{1}{2}},1,0}=-\frac{\sqrt{6}}{2}a_0ieE
\end{equation}
\begin{equation}\label{v45}
\bra{2P_{\frac{1}{2}},1,1}e \vec{E} \cdot \vec{r}\ket{2S_{\frac{1}{2}},0,0}=\frac{\sqrt{6}}{2}a_0ieE
\end{equation}
\begin{equation}\label{v47}
\bra{2P_{\frac{1}{2}},1,1}e \vec{E} \cdot \vec{r}\ket{2S_{\frac{1}{2}},1,0}=\frac{\sqrt{6}}{2}a_0ieE
\end{equation}
\begin{equation}\label{v18}
\bra{2P_{\frac{1}{2}},0,0}e \vec{E} \cdot \vec{r}\ket{2S_{\frac{1}{2}},1,1}=-\frac{\sqrt{6}}{2}a_0ieE
\end{equation}
\begin{equation}\label{v38}
\bra{2P_{\frac{1}{2}},1,0}e \vec{E} \cdot \vec{r}\ket{2S_{\frac{1}{2}},1,1}=-\frac{\sqrt{6}}{2}a_0ieE
\end{equation}

From those results and from Eqs.~(\ref{1}-\ref{8}) it is possible to see that the state $\ket{6}$ only couples with states $\ket{1}$ and $\ket{3}$. To calculate $\bra{1}e \vec{E} \cdot \vec{r}\ket{6}$ and $\bra{3}e \vec{E} \cdot \vec{r}\ket{6}$, for example, we do:
{\fontsize{9.5}{13}\selectfont
$$
\begin{aligned}
\bra{1}e \vec{E} \cdot \vec{r}\ket{6} &= {} m_1(B)\bra{2P_{\frac{1}{2}},0,0}e \vec{E} \cdot \vec{r}\ket{2S_{\frac{1}{2}},1,-1}  \\
       &+ n_1(B)\bra{2P_{\frac{1}{2}},1,0}e \vec{E} \cdot \vec{r}\ket{2S_{\frac{1}{2}},1,-1} \\
       &= (n_1-m_1)\frac{\sqrt{6}}{2}a_0ieE
\end{aligned}
$$
}
{\fontsize{9.5}{13}\selectfont
and
$$
\begin{aligned}
\bra{3}e \vec{E} \cdot \vec{r}\ket{6} &= {} m_3(B)\bra{2P_{\frac{1}{2}},0,0}e \vec{E} \cdot \vec{r}\ket{2S_{\frac{1}{2}},1,-1}  \\
       &+ n_3(B)\bra{2P_{\frac{1}{2}},1,0}e \vec{E} \cdot \vec{r}\ket{2S_{\frac{1}{2}},1,-1}\\
       &= (n_3-m_3)\frac{\sqrt{6}}{2}a_0ieE
\end{aligned}
$$
}

\begin{figure}[h!]
\includegraphics[scale=.51]{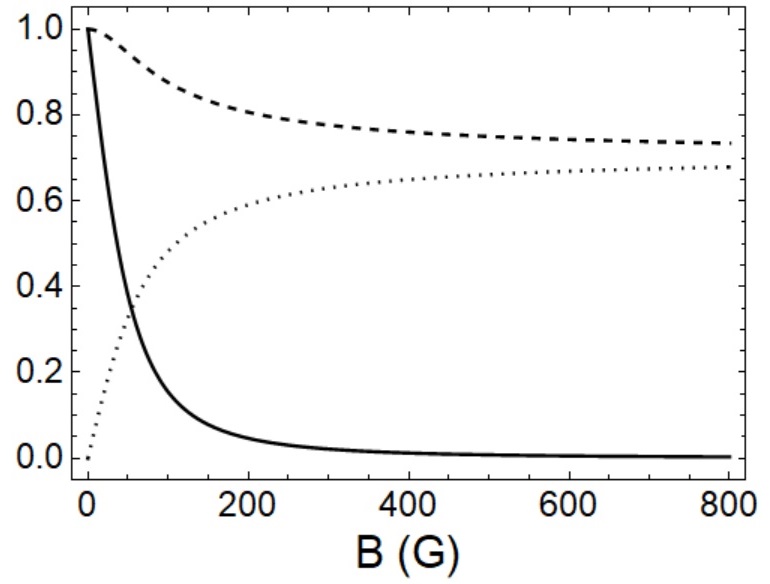}
\caption{The dashed line represents the coefficient $m_1(B)$, the dotted line represents the coefficient $n_1(B)$ and the solid line represents $|n_1(B)-m_1(B)|^2$.}
\label{Fig5}
\end{figure}

\begin{figure}[h!]
\includegraphics[scale=.53]{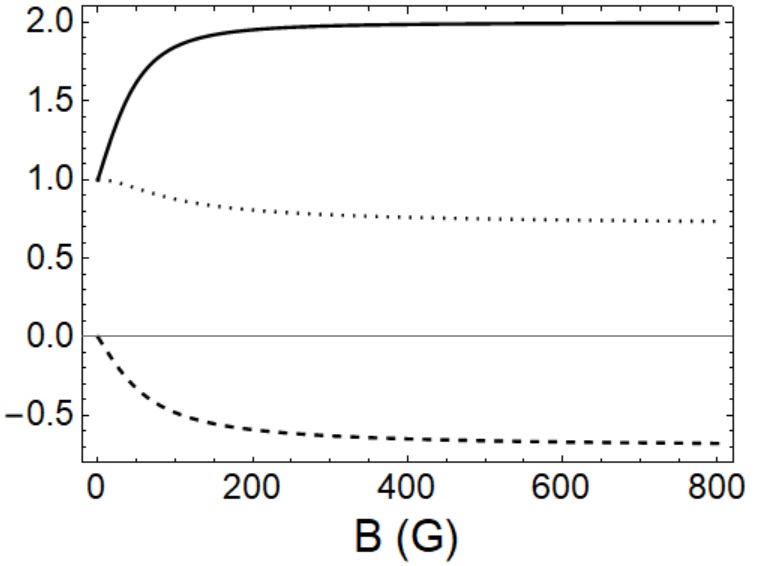}
\caption{The dashed line represents the coefficient $m_3(B)$, the dotted line represents the coefficient $n_3(B)$ and the solid line represents $|n_3(B)-m_3(B)|^2$.}
\label{Fig6}
\end{figure}

Although the analysis of how the probability of the atom being in state $\ket{6}$ changes with time seems to depend on its interaction with states $\ket{1}$ and $\ket{3}$, the interaction of state $\ket{6}$ with state $\ket{1}$ has very little effect on its lifetime. That's due to two main reasons. The first one is that state $\ket{3}$ is closer to state $\ket{6}$, concerning to energy, for magnetic field intensities lower than 800 G and, therefore, has stronger influence in its lifetime. But even more important is the fact that $|V_{16}|^2=(n_1-m_1)^23a_{0}^2e^2E^2/2$ is negligible when compared to $|V_{36}|^2=(n_3-m_3)^23a_{0}^2e^2E^2/2$ for most magnetic field intensities. That happens because $(n_1-m_1)^2$ goes fast to zero with the increase of the magnetic field intensity while $(n_3-m_3)^2$ goes to two, as shown in Figs.~\ref{Fig5} and \ref{Fig6}. 

In order to plot $|c_6(t)|^2$, Figs.~\ref{Fig2}-\ref{Fig4}, we have calculated $\gamma_{+}=(\gamma_{2s}+\gamma_{2p})/2 \pm i\omega_{36}$ and $\omega_{0_+}^2= |V_{36}|^2/\hbar^2+\gamma_{2s}\gamma_{2p}/4 + i\gamma_{2s}\omega_{36}/2$, as mentioned previously.

Determining $\omega_{36}$ is even simpler than determining $|V_{36}|^2$, since we already know how the energies of the states of interest vary with the magnetic field.

\begin{equation}
\omega_{36}(B)=\frac{h_3(B)-h_6(B)}{\hbar}
\end{equation}

Similarly to what was done for state $\ket{6}$, we can study the probability of the atom being in any of the Zeeman states by considering its coupling, by the electric field, with only one other state. Each state $\ket{j}$ couples with only two states $\ket{k}$ and one of these coupling is much stronger than the other.

The determination of $|V_{kj}|^2$ and $\omega_{kj}$ are all what is necessary, taking that we already know $\gamma_{2s}$ and $\gamma_{2p}$, to calculate $\gamma_{\pm}$ and $\omega_{0_{\pm}}^2$ and, therefore, to study the probability of the atom being in any of the states (\ref{1}-\ref{8}), given that it occupies the corresponding $2S_{1/2}$ Zeeman state at $t=0$.

\section{Effects of magnetic field intensity and atomic velocity}
\label{apendD}

\begin{figure}[h!]
\includegraphics[scale=.49]{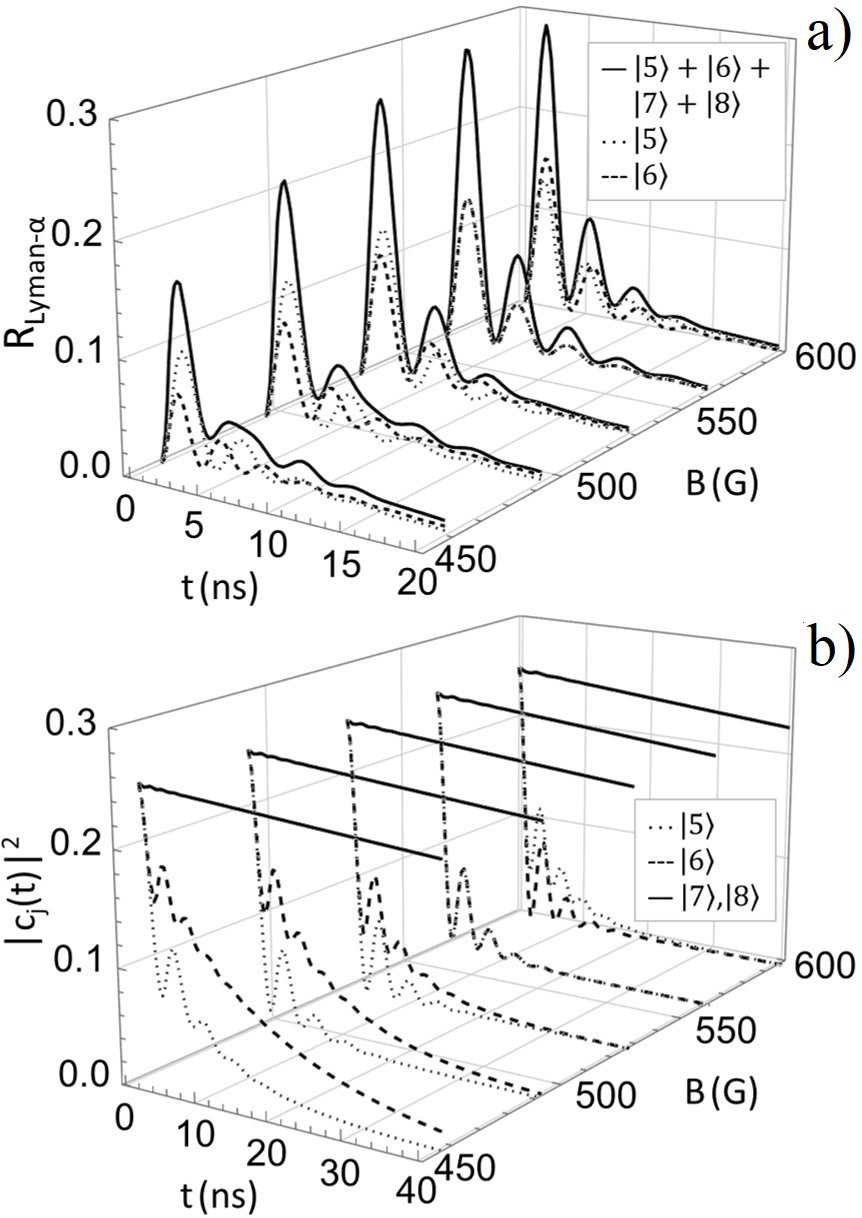}
\caption{a) Represents the Lyman-$\alpha$ radiation rate of an H(2S) beam travelling through a magnetic field region, with velocity 100 km/s perpendicular to the field, for five different values of magnetic field intensity: 460 G, 495 G, 530 G, 565 G, 600 G. The dotted and dashed lines represent the contribution from atoms originally in state $\ket{5}$ and $\ket{6}$, respectively. The solid line represents the Lyman-$\alpha$ radiation rate from all four states. Contributions from states $\ket{7}$ and $\ket{8}$ are not represented here due to their negligible influence on the total emission. b) Represents the populations of states $\ket{5}$, $\ket{6}$, $\ket{7}$ and $\ket{8}$ for the same atomic velocity and magnetic field intensities represented in a). The distinction between the populations of states $\ket{7}$ and $\ket{8}$ are not intended to be perceived from the image, it is only important to observe that they suffer little loss.}
\label{Fig7}
\end{figure}

As mentioned in Section~\ref{sec:fast}, the magnetic field of the polarizer is capable of preserving states $\ket{7}$ and $\ket{8}$ while completely quenching states $\ket{5}$ and $\ket{6}$. This is true for a wide range of magnetic field intensities as can be seen in Fig.~\ref{Fig7}.
We can see that the desired filtering is possible for all considered magnetic field intensities.

Fig.~\ref{Fig7}a) allows us to see that, as mentioned, the Lyman-$\alpha$ emissions from states $\ket{5}$ and $\ket{6}$ get in phase for the magnetic field of 565 G allowing a clearer pattern. 
Fig.~\ref{Fig7}b) shows how the populations of the $\ket{j}$ states change with time for different magnetic field intensities, corresponding to the ones in Fig.~\ref{Fig7}a). 
We have chosen to represent the populations up to 40 ns (corresponding to 4 mm for the chosen atomic speed) because it better represents the selective filtering of the polarizer and helps distinguish between a low rate of Lyman-$\alpha$ emission and a low associated population.
From Fig.~\ref{Fig7}b) we can notice that the decrease in the population happens in a significantly different rate for states $\ket{7}$ and $\ket{8}$ than it does for states $\ket{5}$ and $\ket{6}$. This difference is what allows the polarizer to work. The mentioned difference in decay rate happens for a wide range of magnetic field intensities. For a few centimetres long polarizer it is possible to use a much weaker magnetic field. However, as the field gets weaker, the polarization gets less efficient, meaning that less of states $\ket{7}$ and $\ket{8}$ remain when states $\ket{5}$ and $\ket{6}$ are completely extinct. 
It is important to adjust adequately the magnetic field intensity in the polarizer and its length so that we guarantee that states  $\ket{5}$ and $\ket{6}$ are completely quenched. That adjustment must be done considering experimental convenience.

\begin{figure}[h!]
\includegraphics[scale=.49]{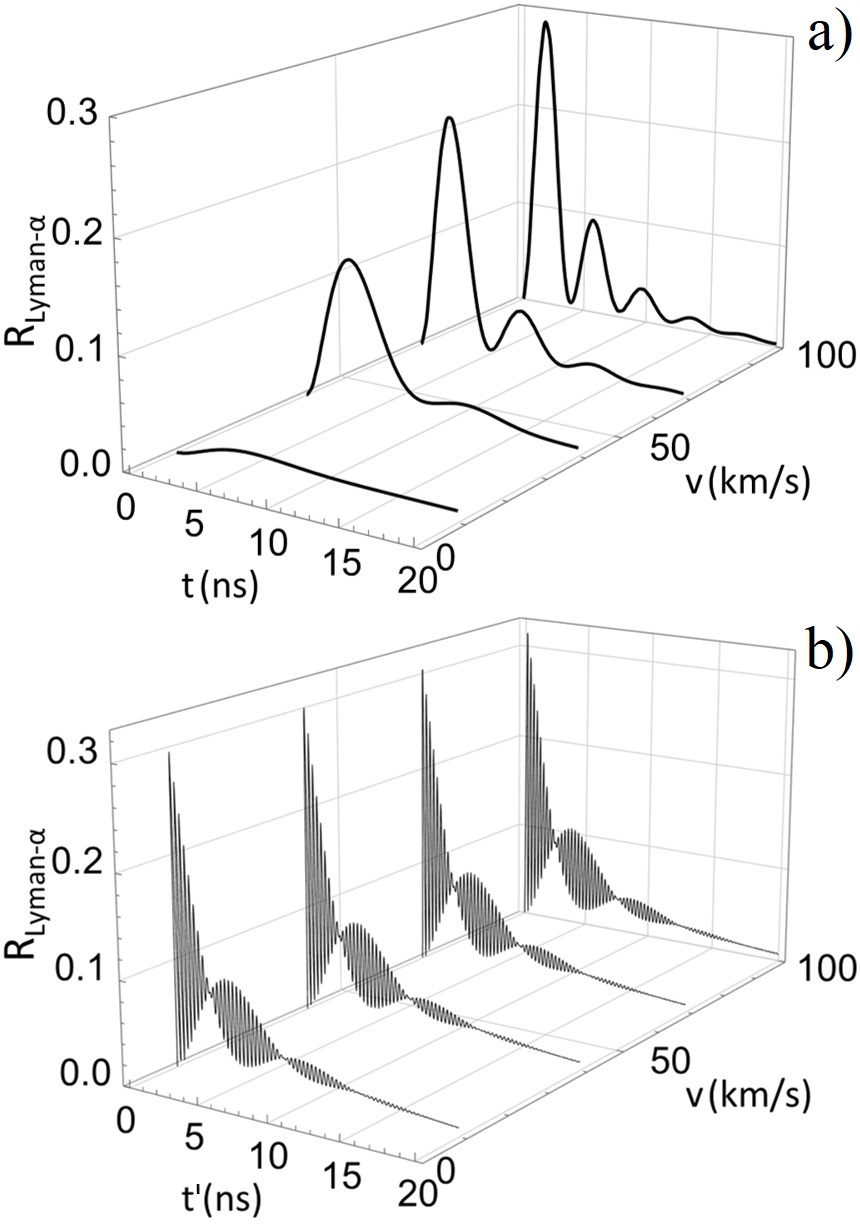}
\caption{a) Lyman-$\alpha$ radiation rate from the polarizer, with a perpendicular magnetic field of 565 G, for four atomic velocities: 10 km/s, 40 km/s, 70 km/s and 100 km/s. b) Lyman-$\alpha$ radiation rate from quenching of states $\ket{7}$ and $\ket{8}$ in the presence of a magnetic field of 565 G and an external electric field of 600 V/m for the velocities: 10 km/s, 40 km/s, 70 km/s and 100 km/s. Here, $\vec{v}$, $\vec{B}$ and $\vec{E}_{ext}$ are perpendicular to each other.}
\label{Fig8}
\end{figure}

The polarizer and its testing criterion are also robust to variations in the atomic velocity. As mentioned previously, the pattern from detector B in Fig.~\ref{Fig4}a) is much more robust to changes in atomic velocity than the patten from detector A. That difference can be noticed from Fig.~\ref{Fig8}.
However, the polarizer's operation is robust to wide variations in beam velocity, including a wide velocity distributions, despite the pattern from detector A being easily lost.

The effects of magnetic field intensity and velocity changes shown in Fig.~\ref{Fig7}a) and Fig.~\ref{Fig8}a) are somewhat related but not equivalent. Although they both influence the electric field in the atom frame, the magnetic field has an important effect on states energies that ultimately allows the filtering.

Fig.~\ref{Fig8}a) shows, for atomic velocities varying from 10 km/s to 100 km/s, the Lyman-$\alpha$ radiation rate emitted from a non-polarized H(2S) beam propagating perpendicular to a magnetic field of 565 G. The patter of emission changes with atomic velocity, but the desired polarization should occur if the beam propagates long enough through the field region. 
From Fig.~\ref{Fig8}b), we see that the proposed pattern for validating the polarizer's effectiveness, for $\vec{E}_{ext}$ = 600 V/m, remains almost unchanged for atomic velocities varying from 10 km/s to 100 km/s. This means it can be used to validate the polarizer's effectiveness for a wide range of atomic velocities. However, this pattern may not be observed, even for a successful polarization, if the beam either has a wide velocity distributions or has to travel a long way to the detection region, because those two factors affect the moment of detection. 

From the understanding of how atomic velocities and field intensities affect the lifetimes of our states of interest, we can determine the relations between atomic velocity, magnetic field intensity and length of the field region required for a two-state polarizer as well as a criterion for validating its effectiveness.

\bibliographystyle{unsrt}
\bibliography{ReF}

\end{document}